\documentclass[twocolumn,pra,aps,superscriptaddress,longbibliography]{revtex4-2}
\usepackage{amsmath}
\usepackage{amssymb}
\usepackage{amsfonts}
\usepackage[dvips]{graphicx}
\usepackage{subfigure}
\usepackage{dcolumn}
\usepackage{txfonts}
\usepackage{bm}
\usepackage{makeidx}
\usepackage{color}
\usepackage{mathtools}
\usepackage{threeparttable}
\usepackage{physics}
\usepackage[colorlinks,linkcolor=blue,anchorcolor=blue,citecolor=blue,urlcolor=blue]{hyperref}
\usepackage{lipsum}

\begin{document}
\title{Limits of single-photon storage in a single $\Lambda$-type atom}

 \author{Zhi-Lei Zhang}
 \affiliation{Center for Quantum Sciences and School of Physics, Northeast Normal University, Changchun 130024, China}
 \affiliation{Graduate School of China Academy of Engineering Physics, Beijing 100193, China}
 
 \author{Li-Ping Yang}
 \email{lipingyang87@gmail.com}
 \affiliation{Center for Quantum Sciences and School of Physics, Northeast Normal University, Changchun 130024, China}

\begin{abstract}
We investigate theoretically the limits of single-photon storage in a single $\Lambda$-type atom, specifically the trade-off between storage efficiency and storage speed. We show that a control field can accelerate the storage process without degrading efficiency too much. However, the storage speed is ultimately limited by the total decay rate of the excited state involved. For a single-photon pulse propagating in a regular one-dimensional waveguide, the storage efficiency has an upper limit of $50 \%$. Perfect single-photon storage can be achieved by using a chiral waveguide or the Sagnac interferometry. By comparing the storage efficiencies of Fock-state and coherent-state pulses, we reveal the influence of quantum statistics of light on photon storage at the single-photon level.
\end{abstract}

\maketitle

\section{Introduction}
Quantum memories for photon pulses are crucial for quantum communications~\cite{duan2001long,kimble2008quantum,simon2017towards} and quantum computing~\cite{kok2007linear,cirac1999distributed}. Via the photon echo technique or electromagnetically induced transparency (EIT) effect, storage of weak coherent-state pulses with efficiency $\approx 90\%$ has been achieved~\cite{hedges2010efficient,cho2016highly,geng2014electromagnetically,Chen2013coherent,vernaz2018highly,Hsiao2018highly}. Recently, the storage of Fock-state single-photon (FSSP) pulses with efficiency $>85\%$ has also been realized with laser-cooled rubidium atoms~\cite{wang2019efficient}. However, in these experiments, the length of the target pulse ($\tau_p$) is around a few to tens of microseconds and it is almost three orders of magnitude larger than the lifetime $1/\gamma$ of the involved excited state of atoms. High-speed optical quantum memories for short pulses ($\approx 1$~ns) has also been demonstrated~\cite{reim2010towards}, but the storage efficiency is relatively low ($< 30\%$)~\cite{corzo2019waveguide}. Storage of single-photon pulses with high efficiency and high speed remains a challenge~\cite{Heshami2016Quantum,Ma2017Optical,Shi2018Quantum}. 

Compared with an atomic ensemble~\cite{Fleischhauer2000dark,Fleischhauer2002quantum,Kozhekin,Gorshkov2007universal,Nunn2007Mapping}, single-atom system~\cite{Boozer2007Reversible,Specht2011,giannelli2018optimal,Meng2020imaging} provides a novel platform to explore the fundamental limits of single-photon storage, specifically, the trade-off between storage efficiency and storage speed. A closely related problem, i.e., single- or few-photon scattering by an atom, has been extensively studied~\cite{shen2005coherent,Zhou2008Controllable,shi2009Lehmann,witthaut2010photon,roy2011twophoton,zheng2013Persistent,zhou2013quantum}. Recently, the time-delay induced interference effect attracts new interests about photon scattering by a giant atom ~\cite{guo2017giant,zhao2020single,gu2017microwave,kockum2018,guo2020oscillating,wang2021tnable,zou2022tunable,yin2022single}. However, these research works focus more on the reflection and transmission coefficients, not figures of merit of storage. On the other hand, the impact of photon number quantum fluctuations, which plays a crucial role in light-atom interaction at the single-photon level, has not been adequately explored. 

In this work, we investigate the limits of single-atom-based  single-photon storage without and with a control field. For a three-level atom placed in a regular one-dimensional waveguide, there exists an upper limit ($50 \%$) on the single-photon storage efficiency. A chiral waveguide~\cite{Gonzalez2016Nonreciprocal,Du2021nonreciprocal,wang2022unconventional,Pucher2022atomic} or Sagnac interference technique~\cite{Shen2012efficient,shen2012-single,Du2021single} could be used to improve the efficiency and to realize perfect storage. In the absence of a control field, we find that high storage efficiency could be obtained only for long single-photon pulses ($\tau_p\gg 1/\gamma$). Thus, there is a trade-off between storage efficiency and storage speed. A control field could be applied to enhance the storage speed and improve the storage efficiency for single-photon pulses with length $\tau_p=1/\gamma$. However, the storage speed is ultimately limited by the total decay rate of the involved excited state. Different from an atomic ensemble, a single multilevel atom exhibits high nonlinearity. We show that the storage efficiency of a coherent-state single-photon (CSSP) pulse is lower than that of a FSSP pulse, since nonlinear multiphoton processes have been suppressed.

This article is structured as follows: In Sec.~\ref{sec:masterEQ}, we begin by introducing the master equation for a single $\Lambda$-type atom driven by a quantum pulse. In Sec.~\ref{sec:nocontrol}, we investigate the storage of single-photon pulses without a control field. In Sec.~\ref{sec:control}, we show the storage speed could be accelerated via a control field. In Sec.~\ref{sec:PerfectStorage}, we show a chiral waveguide and the Sagnac interferometer could be exploited to realize perfect storage of FSSP pulse. We brieﬂy summarize in Sec.~\ref{sec:summary}. Some details about the master equation are given in the Appendix%~\ref{sec:appendix}.

\begin{figure}
    \centering
    \includegraphics[width=8.5cm]{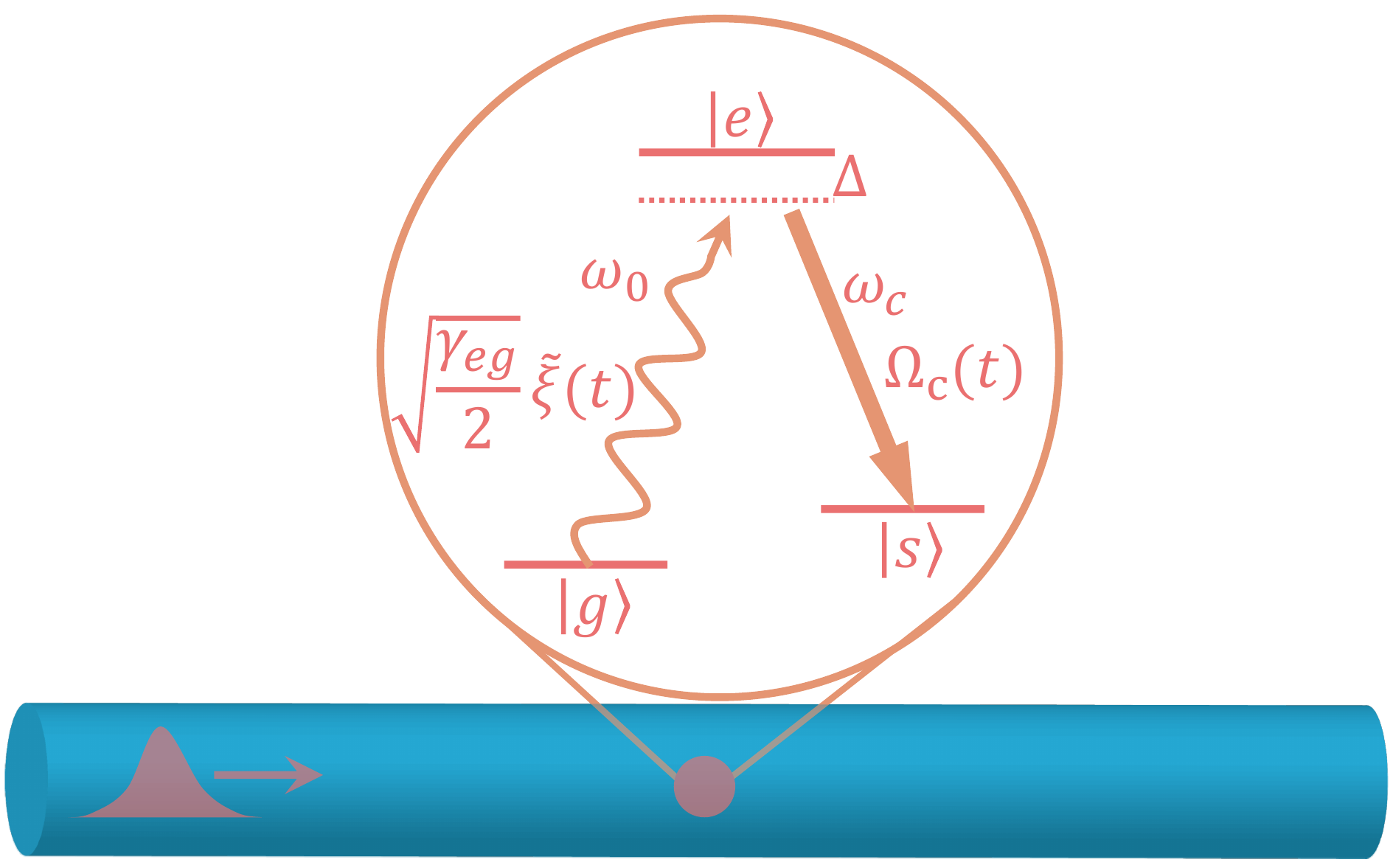}   \caption{\label{three_levels_atom}Scattering of a single-photon pulse with center frequency $\omega_0$ and wave-packet function $\tilde{\xi}(t)$ by a three-level $\Lambda$-type atom placed in a one-dimensional waveguide. A control pulse with frequency $\omega_c$ and strength $\Omega_c (t)$ is applied to assist the storage process. The decay rates of the excited state $|e\rangle$ to the two ground states are $\gamma_{eg}$ and $\gamma_{es}$ and $\Delta$ is the two-photon detuning. }
\end{figure}

\section{Master equations for a $\Lambda$-type atom driven by a quantum pulse\label{sec:masterEQ}}

Recently, substantial efforts have been devoted to investigating the scattering of propagating quantum pulses by a local quantum system~\cite{Macha2020,giannelli2018optimal,Shi2015multiphoton,Liao2015single-photon,Liao2016Dynamical}. A systematic master-equation approach has been developed to handle the dynamics the local quantum scatter~\cite{gheri1998photon,combes2012Nphoton,combes2017slh}. The input-output relation has also been incorporated to give the information of the outgoing temporal mode~\cite{Kiilerich2019input,Kiilerich2020quantum}. Here, we follow the approach given in Ref.~\cite{gheri1998photon} to handle the storage of both FSSP and CSSP pulses in a single $\Lambda$-type atom. We show that the quantum statistics of the quantum pulse affect the storage efficiency significantly. 

The basic elements of the storage process are illustrated in Fig.~\ref{three_levels_atom}. The $\Lambda$-type atom, which is described by Hamiltonian $H_a = \omega_g |g\rangle\langle g| + \omega_e |e\rangle\langle e|+\omega_s |s\rangle\langle s|$, contains two stable ground states $|g\rangle$ and $|s\rangle$ and one excited state $|e\rangle$. For a regular one-dimensional waveguide, both the forward-propagating modes $a(\omega)$ and backward-propagating modes $b(\omega)$  have to be considered. The Hamiltonian for the waveguide photons is given by $H_p = \int d\omega (\omega_0+\omega) [a^{\dagger}(\omega) a(\omega) + b^{\dagger}(\omega) b(\omega)]$, where the frequency of the waveguide photons has been expanded to the first-order of the wave-vector along the propagating direction around the near-resonant mode  $\omega_0$~\cite{shen2009theory,huang2013controlling}. The interaction between the atom and waveguide photons is described by 
\begin{equation}
 H_{\rm int}\!=\!\!\int\!\! d\omega\! \left[g_{eg}(\omega) \sigma_{ge}^{\dagger}\! +\! g_{es}(\omega) \sigma_{se}^{\dagger}\right] [a(\omega)\! +\! b(\omega)]\! +\! {\rm H.C.}, \label{eq:Hint}
\end{equation}
where $\sigma_{ge}=|g\rangle\langle e|$, and $\sigma_{se}=|s\rangle\langle e|$. In addition, an extra control laser pulse could be applied to assist and accelerate the storage process. The interaction to the control field is described by Hamiltonian $H_c =  [\Omega_c(t)\exp(-i\omega_c t)\sigma_{se}+\rm{H.C.}]$. To enhance the storage efficiency, the two-photon-resonance condition is required ,i.e., $\omega_e-\omega_0 = \omega_e-\omega_s-\omega_c = \Delta$.

Both CSSP pulses and FSSP pulses have been commonly used in storage experiments~\cite{liu2022ondemand,Jelena2022Storage}. The dynamics of a nonlinear scatter exhibit very different features under these two types of quantum pulses~\cite{gheri1998photon,Wang2011Efficient,yang2018concept}. The single-photon wave-packet creation operator $a_{\xi}^{\dagger}=\int d\omega \xi(\omega)a^{\dagger}(\omega)$ is usually used to generate quantum photon pulse wave function~\cite{loudon2000quantum}. The pulse shape is determined by the normalized spectral amplitude function $\int d\omega |\xi(\omega)|^2=1$. Forward-propagating CSSP and FSSP pulses are described by $|1_{\rm CS}\rangle = \exp(a^{\dagger}_{\xi}-1/2)|0\rangle$ and $|1_{\rm FS}\rangle = a^{\dagger}_{\xi}|0\rangle$, respectively. Initially, the atom is prepared in the ground state $|g\rangle$. The incident single-photon quantum pulse excites the atom and  transfers it to state $|s\rangle$ to realize the storage.

A CSSP pulse can be treated as a classical driving field. The dynamics of the atom density matrix are governed by a Lindblad master equation $\dot{\rho}(t) = [\mathcal{L}_{ac}+\mathcal{L}_p(t)]\rho(t)$, where
\begin{align}
\mathcal{L}_{ac} \rho(t) = &- i [H_a+H_c,\rho(t)] - \frac{\gamma_{eg}+\gamma_{es}}{2} \{ |e\rangle\langle e| , \rho(t) \}  \nonumber\\ 
& + \gamma_{eg}\sigma_{ge} \rho(t) \sigma_{ge}^{\dagger}+\gamma_{es} \sigma_{ge} \rho(t) \sigma_{ge}^{\dagger},
\end{align}
describes the spontaneous decay of the excited state $|e\rangle$ with a classical control on the storage channel. The pumping of the atom by the CSSP pulse is described by the Liouville operator~\cite{gheri1998photon}
\begin{equation}
\mathcal{L}_p (t)\rho(t) =  -i\sqrt{\frac{\gamma_{eg}}{2}} \left\{ \left[\tilde{\xi}(t)\sigma_{ge}^{\dagger}, \rho(t)\right]+  \left[\tilde{\xi}^{*}(t)\sigma_{ge}, \rho^{\dagger}(t)\right]\right\},\label{eq:Lp}
\end{equation}
where $\tilde{\xi}(t)$ is the wave-packet function of the CSSP pulse determined by the Fourier transform of $\xi(\omega)$~\cite{yang2018concept}. We emphasize that there is a factor $1/\sqrt{2}$ in $\mathcal{L}_p$, because both the forward and backward waveguide modes will contribute to the decay of the excited state $|e\rangle$, but the target pulse only contains forward modes. Here, we see that a CSSP pulse functions as a classical driving, since $\rho^{\dagger}(t)=\rho(t)$.

The traditional Lindblad master equation cannot be used to describe the interaction between a FSSP pulse and a localized quantum system~\cite{gheri1998photon}. A generalized Fock-state master equation has been developed~\cite{gheri1998photon,combes2012Nphoton}, 
\begin{align}
    \dot{\rho}(t) &= \mathcal{L}_{ac} \rho(t) +\mathcal{L}_p(t)\rho_{01}(t) \label{evolution equation of rhoS}\\
    \dot{\rho}_{01}(t) &= \mathcal{L}_{ac} \rho_{01}(t) - i \sqrt{\frac{\gamma_{eg}}{2}} \tilde{\xi}^*(t) [\sigma_{ge},\rho_{00}(t)], \label{eq:MEQrho_01}\\
    \dot{\rho}_{00} (t) &= \mathcal{L}_{ac} \rho_{00}(t), \label{evolution equation of rhor}
\end{align}
where
\begin{equation}
\begin{aligned}
\rho(t) &= {\rm Tr}_R [U(t) \rho(0) \otimes |1_{\rm FS}\rangle\langle 1_{\rm FS} |\otimes |0_b\rangle\langle 0_b| U^{\dagger}(t)],\\
\rho_{01}(t) &= {\rm Tr}_R[U(t) \rho(0) \otimes |0_a\rangle\langle 1_{\rm FS}|\otimes |0_b\rangle\langle 0_b| U^{\dagger}(t)],\\
\rho_{00}(t) & = {\rm Tr}_R[U(t) \rho(0) \otimes |0_a\rangle\langle 0_a|\otimes |0_b\rangle\langle 0_b| U^{\dagger}(t)],
\end{aligned}
\end{equation}
and $U(t)=\mathcal{T} \exp[-i \int_0^t (H_a + H_p +H_c +H_{\rm int}) dt]$ is the time evolution operator of the whole system. The initial-state waveguide modes is $|1_{\rm FS}\rangle\otimes |0_{b}\rangle$. We note that significantly different from a CSSP pule, the pumping by a FSSP pulse [i.e., $\mathcal{L}_p(t)\rho_{01}(t)$] can not be regarded as a classical driving since $\rho_{01}^{\dagger}(t)\neq \rho_{01}(t)$.

\begin{figure}
    \centering   \includegraphics[width=8.5cm]{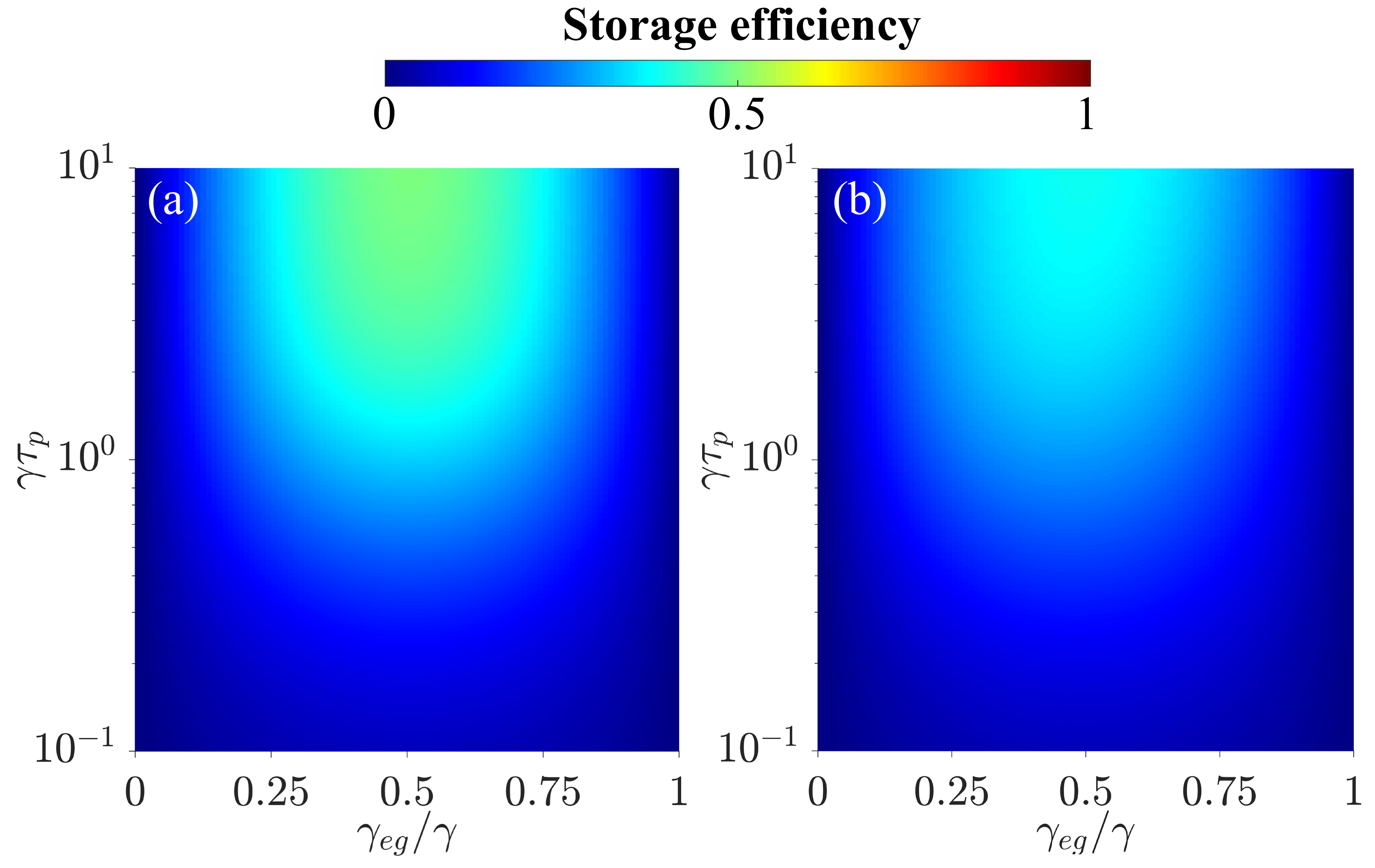}
    \caption{\label{fig2}Optimization of the storage efficiency for (a) a Fock-state single-photon pulse and (b) a coherent-state single-photon by varying pulse length $\tau_p$ and decay rate $\gamma_{eg}$. No control field is applied (i.e., $\Omega_c = 0$) and the two-photon detuning $\Delta$ is set as zero.}
\end{figure}

\begin{figure}[!htbp]
    \centering
    \includegraphics[width=8.8cm]{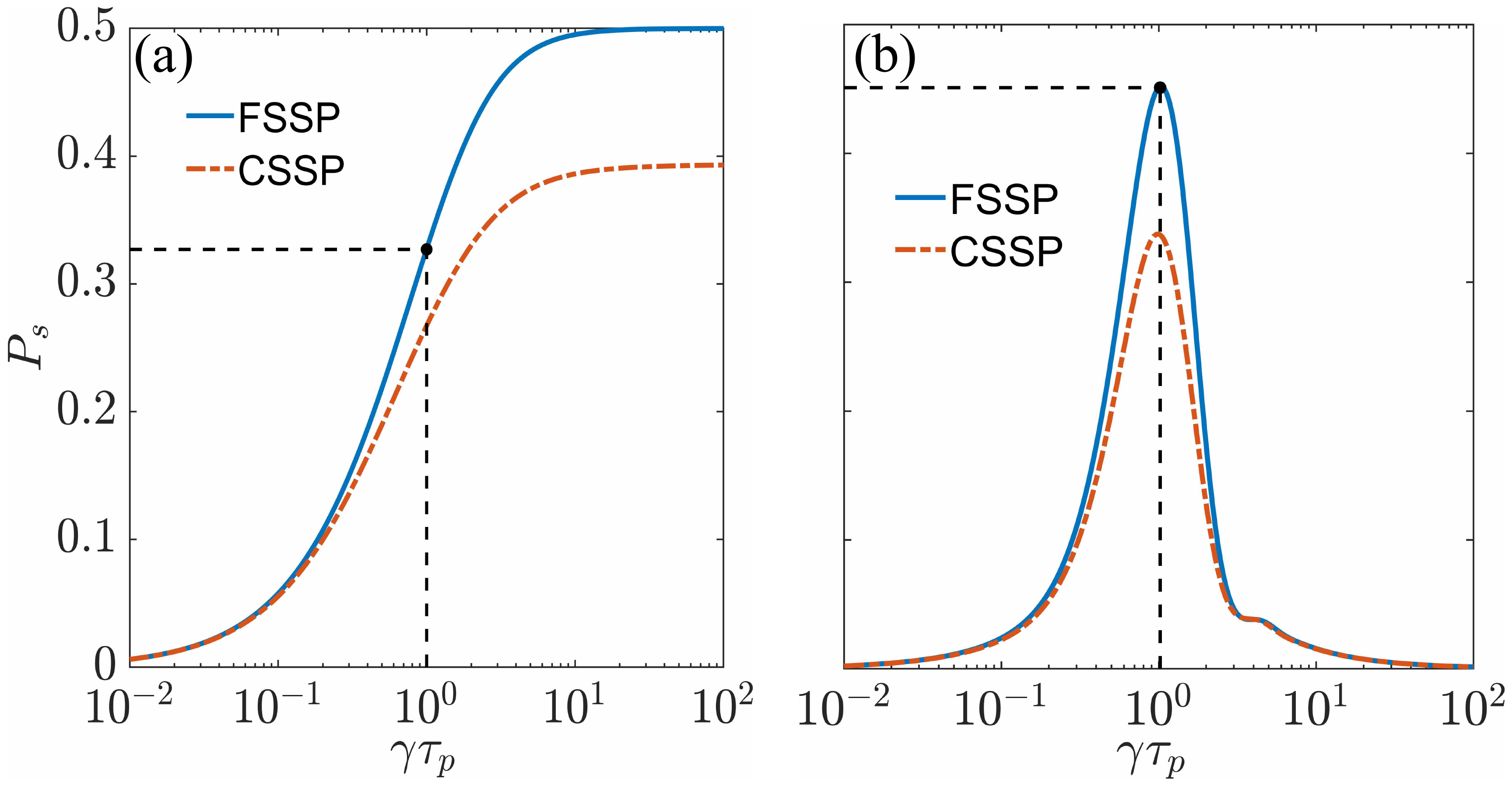}
    \caption{\label{fig3}Contrast between the storage efficiency of a Fock-state single pulse (FSSP) and a coherent-state single-photon (CSSP) pulse. The two-photon detuning $\Delta$ is set as zero. (a) Optimized storage efficiency in the absence of control field with $\Omega_c = 0$ and $\gamma_{eg} = \gamma_{es} = \gamma/2$. (b) Optimized storage efficiency in the presence of a control field with strength $\Omega = 0.7 \gamma$, length $a = 0.9\tau_p$, and relative delay $b = 0.6\tau_p$. The other parameters have been taken as $\gamma_{eg} = 0.9 \gamma$, and $\gamma_{es} = 0.1 \gamma$.}
\end{figure}

\section{Storage of a single-photon pulse without control field \label{sec:nocontrol}}

In this section, we study the storage of a single-photon pulse in the absence of a control pulse, i.e., $\Omega_c = 0$. The advantage of this storage scheme is that no information about the arrival time of the target pulse is needed. The atom initially prepared in state $|g\rangle$ will be excited to state $|e\rangle$ and spontaneously decays to state $|g\rangle$ or the storage state $|s\rangle$. The storage efficiency of a single-photon pulse is defined as the steady-state probability $P_s$ of state $|s\rangle$. We show that the decay rates of the storage channel and the pumping channel must be carefully matched to optimize storage efficiency. We also show that the storage efficiency of a CSSP pulse will be lower than that of a FSSP pulse.

There are three parameters to optimize the storage efficiency, i.e., the two decay rates $\gamma_{eg}$ and $\gamma_{es}$ and the length of the target pulse. Without loss of generality, we assume the target pulse is of the Gaussian shape,
\begin{equation}
    \tilde{\xi}(t) = \left(\frac{1}{2 \pi\tau_p^2}\right)^{\frac14} \exp\left[-\frac{\left(t - t_0 \right)^2}{4 \tau_p^2}\right], \label{xi in real space}
\end{equation}
where $t_0 $ is the time of the pulse arriving at the atom and $\tau_p$ is the half-length of the pulse. In the following, we fix the total decay rate $\gamma = \gamma_{eg}+\gamma_{es}$ of state $|e\rangle$ and take it as the unit of frequency, i.e., $\gamma = 1$. It is usually challenging to continuously adjust the decay rates $\gamma_{eg}$ and $\gamma_{es}$ for a natural atom situated in a waveguide. However, in the case of alkali-metal atoms, both the ground and excited states comprise numerous degenerate hyperfine sublevels that can be broken using an external magnetic field. The transition strength between two electronic states is determined not only by the electric-dipole transition matrix elements but also by the overlap between their spin states, i.e., the Clebsch–Gordan coefficients. By carefully selecting the hyperfine states, the parameters required for our study could be obtained.

Maximum storage efficiency will be obtained if the decay rates of the pumping and storage channels are equal to each other, i.e., $\gamma_{eg} = \gamma_{es} = \gamma/2$. In Figs.~\ref{fig2}(a) and~\ref{fig2}(b), we plot the storage efficiencies for a FSSP pulse and a CSSP pulse, respectively, as a function of $\gamma_{eg}$ and pulse length $\tau_p$. For a given pulse length, the maximum storage efficiency locates at $\gamma_{eg} = \gamma /2$ for both FSSP and CSSP pulses. There is a slight asymmetry present in Fig.~\ref{fig2}(b) caused by the stimulated emission induced by the multiphoton components of a CSSP pulse. However, this asymmetry is not present in a long pulse (not shown) since the stimulated radiation can be neglected in such cases. On the other hand, the storage efficiency of a longer pulse is larger. This can be seen more clearly in Fig.~\ref{fig3}(a). To obtain higher storage efficiency, one needs to sacrifice the storage speed.

\begin{figure}
    \centering
    \includegraphics[width=8.8cm]{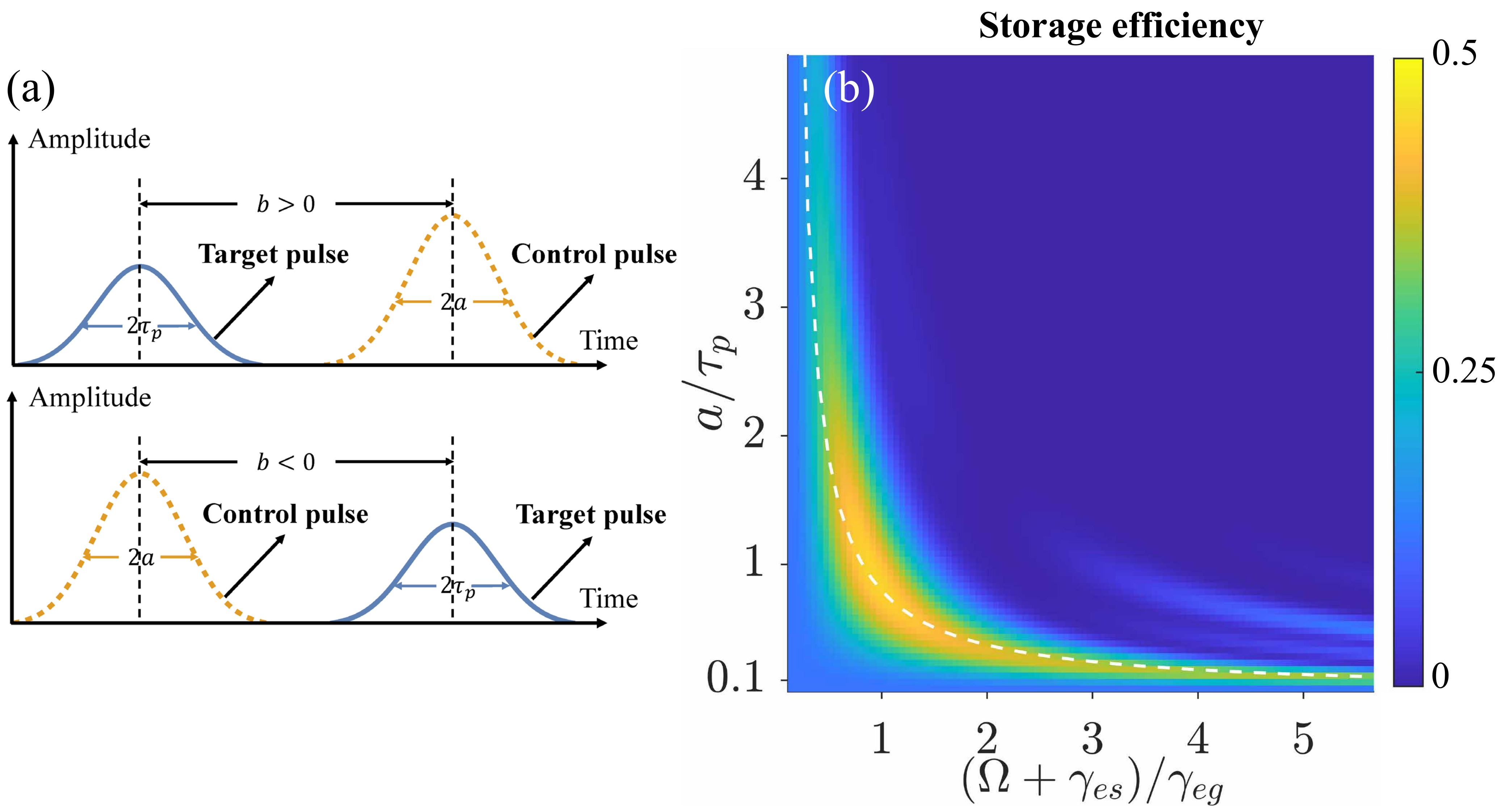}
    \caption{\label{fig4}(a) Sketch map of the relative delay $b$ between the target pulse (blue solid line) and the control pulse (orange dashed line). (b) Storage efficiency of a Fock-state pulse varies with the magnitude $\Omega$ and width  $a$ of a control pulse. $\gamma_{eg} = 0.9\gamma$, $\gamma_{es} = 0.1\gamma$, and $b = 0.6 \tau_p$. The fitting white dashed line  $2a\Omega\sqrt{\pi} = 2.26$ characterizes the constant area under the envelope function $\Omega_c (t)$.}
\end{figure}

\begin{figure*}
    \centering  
    \includegraphics[width=16cm]{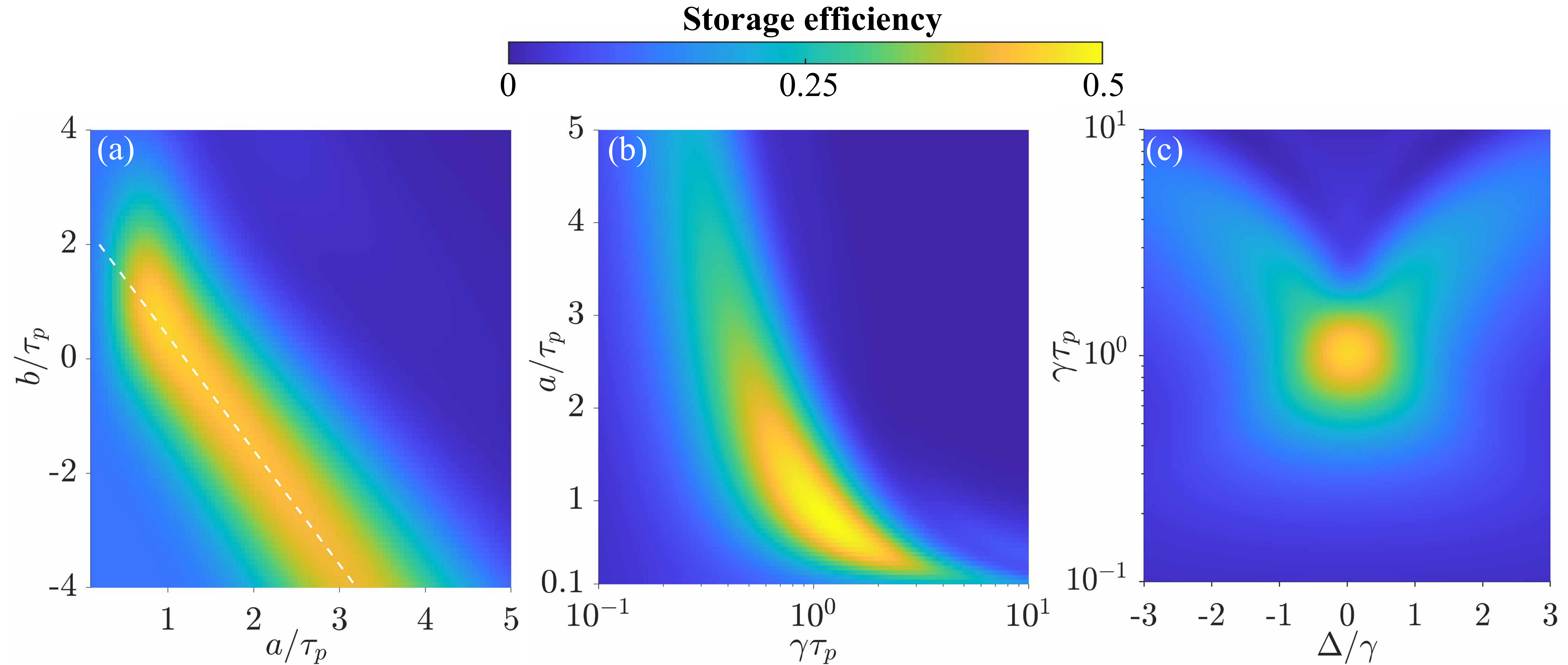}
    \caption{\label{fig5}Optimization of storage efficiency of a Fock-state single-photon pulse with $\gamma_{eg} = 0.9\gamma$, $\gamma_{es} = 0.1\gamma$, and $\Omega= 0.7 \gamma$. (a) Optimization with fixed pulse length $\tau_p = 1/\gamma$ and two-photon detuning $\Delta=0$. The fitting white dashed line is given by $b + 2 a = 1.2\times 2 \tau_p$. (b) Optimization with fixed delay $b = 0.6 \tau_p$ and $\Delta=0$. (c)  Optimization with half-length $ a = 0.9 \tau_p$ and delay $b = 0.6 \tau_p$ of the control pulse.}
\end{figure*}

The storage efficiency is strongly affected by the quantum statistics of the target pulse. As shown in Fig.~\ref{fig3}(a), the storage efficiency of a FSSP pulse is higher than that of a CSSP pulse. A CSSP pulse has a high probability of not containing any photons, resulting in zero excitation. The few-level atom functions as a nonlinear system~\cite{yang2018concept,Yao_2020}, and multiphoton processes are prohibited. Consequently, the probability of exciting a single atom with a CSSP pulse is notably lower compared with a Fock-state single-photon pulse. We also note that there exists an upper limit in the storage efficiency. When $\tau_p\gg 1/\gamma$, the storage efficiency of the FSSP (CSSP) pulse CSSP reaches the upper limit $0.5$ ($0.4$). This low storage efficiency fundamentally results from the fact that the pumping rate is half of the decay rate of the $|g\rangle\leftrightarrow |e\rangle$ channel~\cite{Liao2015single-photon,yang2018concept}. Perfect storage of single-photon pulses can be realized by enhancing the pumping rate as shown in Sec.~\ref{sec:PerfectStorage}.

\section{Storage of a single-photon pulse with a control field \label{sec:control}}
In Sec.~\ref{sec:nocontrol}, we show that the storage efficiency for short single-photon pulses ($\tau_p \leq 1/\gamma$) is relatively low. To assist and accelerate the single-photon storage, an extra control field could be applied to $|e\rangle \rightarrow |s\rangle$ channel~\cite{Fleischhauer2000dark,Kozhekin,Nunn2007Mapping,Boozer2007Reversible}. In the absence of a control pulse, maximum storage efficiency is obtained under the decay-rate matching condition $\gamma_{eg}=\gamma_{es}$. The control pulse provides new parameters, which can be much more easily controlled in experiments, to optimize storage efficiency. We show that the total decay rate $\gamma=\gamma_{eg}+\gamma_{es}$ plays an essential role in the storage process. Specifically, it limits the maximum storage speed. This marks a significant difference from the storage of a single-photon pulse in an atomic ensemble, where more attention was paid to the $\sqrt{N}$-enhanced ($N$ is the atom number) coupling strength between the target photon and the collective atomic states ~\cite{Gorshkov2007I,Gorshkov2007II,Gorshkov2008IV}.

In the following, we take a Gaussian control pulse as an example. Our main results are also valid for other types of control pules. The envelope of the control pulse is given by
\begin{equation}
 \Omega_c(t) = \Omega \exp {- \left(\frac{t - t_0 - b}{2 a}\right)^2}, \label{Gaussian wave packet of control}
\end{equation}
where $\Omega$ characterizes the effective strength of the control pulse, $a$ is its half width, and $t_0$ is the time of the pulse center arriving at the atom. As shown in Fig.~\ref{fig4}(a), $b$ is the relative delay between the target single-photon pulse and the control pulse. In addition to $\gamma_{eg}$ and $\gamma_{es}$, we now have three more easily controlled parameters to optimize single-photon storage.

Similar to the $\Omega_c =0$ case, transition rates between the pumping channel and the storage channel also need to be balanced to obtain larger storage efficiency.  As shown in Fig.~\ref{fig4}(b), maximum storage efficiency locates around $(\Omega +\gamma_{es})/\gamma_{eg}\approx 1$ when the length of the control pulse is long enough. For a short control pulse ($a < \tau_p$), a larger strength $\Omega$ is required to guarantee that the energy of the control pulse is enough to transfer the population from state $|e\rangle$ to state $|s\rangle$. The white dashed line denotes the fitting curve $2a\Omega \sqrt{\pi} = 2.26$, i.e., the area under the envelope function $\Omega_c(t)$ is a constant. To investigate the benefit of the control pulse, we will take $\gamma_{eg} = 0.9\gamma$ and $\gamma_{es}=0.1\gamma$ when a control pulse is applied.

The relative delay $b$ and half-length $a$ of the control pulse need to be matched to obtain larger storage efficiency. In Fig.\ref{fig5}(a), we plot the storage probability $P_s$ of a FSSP pulse as a function of $a$ and $b$. The largest storage efficiency locates at $b= 0.6 \tau_p$ and $a= 0.9 \tau_p$, i.e., a positive delay and a length comparable to the length of the target pulse. A similar delay was also required for an atomic ensemble optical memory~\cite{Nunn2007Mapping}. For a negative delay $b$, higher storage efficiency could also be obtained around the line $b+2a = 1.2\times 2\tau_p$. This guarantees that the control pulse and the target single-photon pulse always have sufficient overlap.

There exists a favorable length $\tau_p$ of the target pulse in storage efficiency optimization with fixed delay $b$ and strength $\Omega$ of the control pulse. A larger storage efficiency could be obtained for $\tau_p= 1/\gamma$ as shown in Fig.~\ref{fig3}(b). This marks a significant difference from the case in the absence of a control pulse, in which longer single-photon pulses ($\tau_p\gg 1/\gamma$) always have higher storage efficiency [see Fig.~\ref{fig3}(a)]. In Fig.~\ref{fig5}(b), we plot the storage probability $P_s$ of a FSSP pulse as a function of $\tau_p$ and $a=0.9\tau_p$ with $b= 0.6 \tau_p$ and $\Omega = 0.7\gamma$. We show that larger storage efficiency is obtained around $\tau_p=1/\gamma$. Thus, the control pulse could be used to improve the storage speed. Previously, off-resonant Raman technique~\cite{Nunn2007Mapping} has been explored to store a single broadband (short) photon in an atomic ensemble beyond the adiabatic storage frame based on EIT~\cite{Fleischhauer2000dark}. However, in the single-atom case, the two-photon detuning $\Delta$ will reduce the storage efficiency greatly, as shown in Fig.~\ref{fig5}(c). Moreover, large storage efficiency is still obtained around $\tau_p=1/\gamma$ for fixed $a$ and $b$. No extra acceleration is obtained with nonzero detuning $\Delta$. The storage of a CSSP pulse is similar to that of a FSSP pulse but with lower efficiency.

Recently, there have been studies on the use of a $\Lambda$-type atom in a cavity for single-photon storage, both theoretically~\cite{giannelli2018optimal} and experimentally~\cite{Macha2020}. However, in practice, there are issues such as leakage to free space and the presence of undesired states. For coupling with an atom, a photon pulse in an ideal one-dimensional waveguide is more efficient than a free-space propagating pulse. As a result, the storage efficiency achieved in Ref.~\cite{Macha2020} was relatively low. Furthermore, we would like to emphasize that our focus in this study is on exploring the limits and trade-offs in single-photon storage, as demonstrated in the following section.

\begin{figure}
    \centering
    \includegraphics[width=8.5cm]{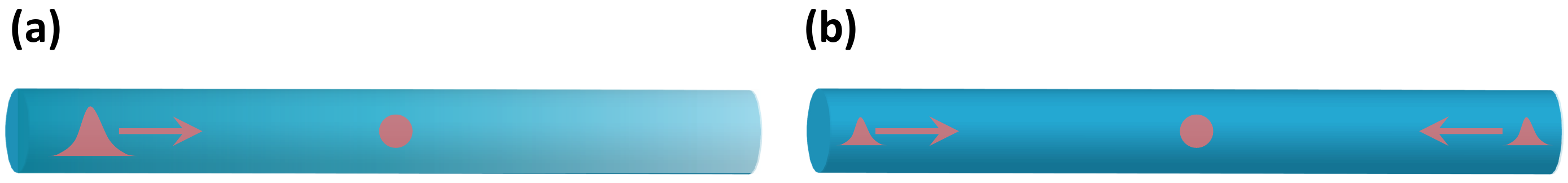}
    \caption{\label{fig:6}Sketch map of two possible approaches to improving storage efficiency: (a) The atom only couples to the forward propagating photons in a perfect chiral waveguide. (b) For the Sagnac interferometry method, the target pulse is split into two smaller pulses, which enter the waveguide at different ends.}
\end{figure}

\begin{figure}
    \centering
    \includegraphics[width=8.5cm]{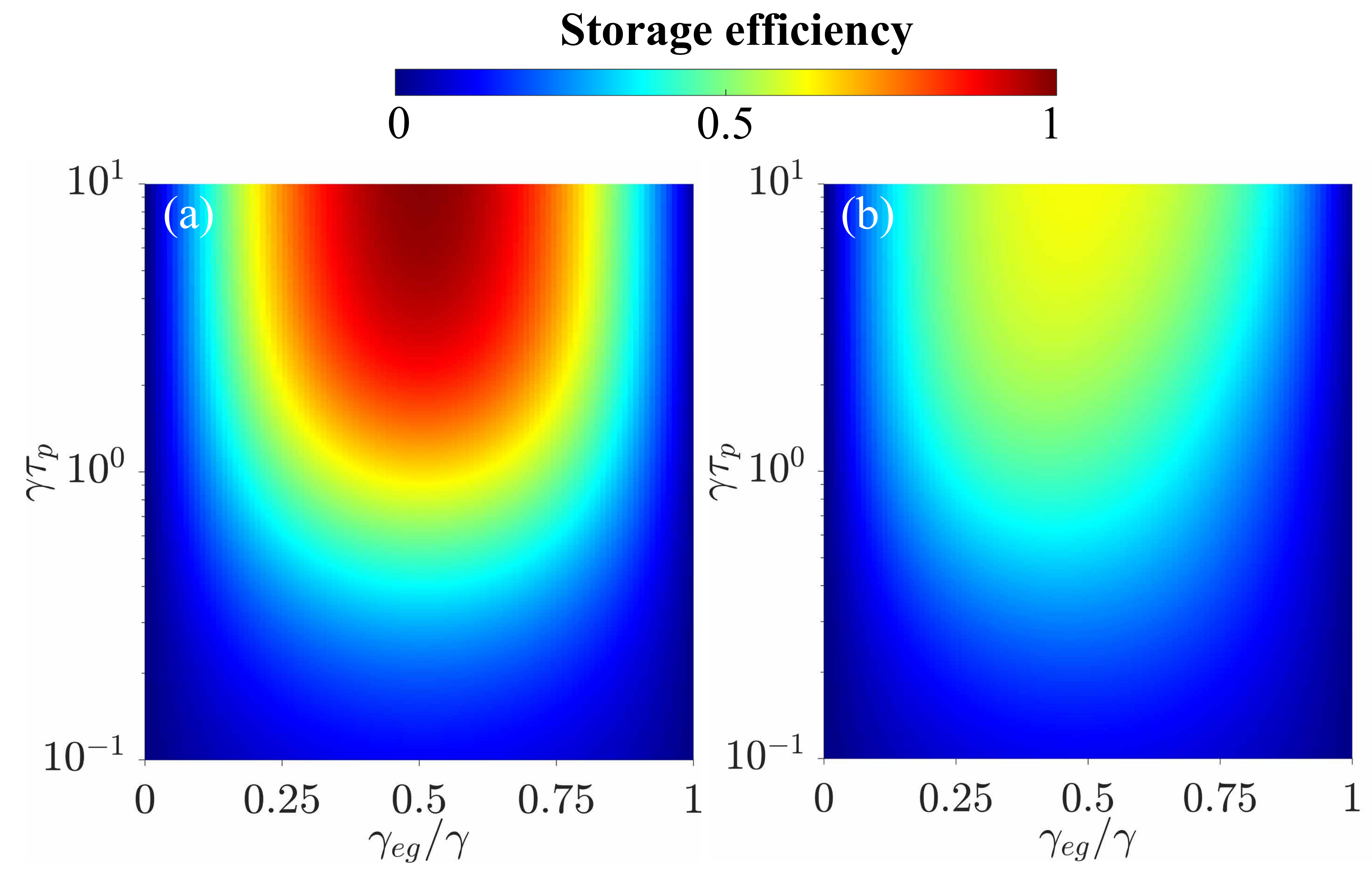}
    \caption{\label{fig:7} Comparison of improved storage efficiency of (a) a Fock-state single-photon pulse and (b) a coherent-state single-photon pulse without a control field.}
\end{figure}

\section{Efficient storage by exploiting a chiral waveguide or a Sagnac interferometer\label{sec:PerfectStorage}}
In previous sections, we show that the storage of a FSSP pulse in a single three-level atom is limited to $0.5$ with or without a control pulse. The storage efficiency for a CSSP is even lower. This low efficiency strongly hampers the practical application of the single-atom storage scheme. In this section, we show that perfect storage of single-photon pulse in a three-level atom can be realized by exploiting a chiral waveguide~\cite{Mitsch2014Quantum,Jan2014chira,Sollner2015Deterministic,Sayrin2015nanophotonic} or a Sagnac interferometer~\cite{Shen2012efficient,shen2012-single}. Previously, these two methods have been applied successfully to enhance the frequency conversion efficiency~\cite{yan2013tunable,jia2017efficient,Du2021single} and to control single-photon transport~\cite{Gonzalez2016Nonreciprocal,Chen2022Nonreciprocal,lu2017coherent,fan2018tunable,Du2021nonreciprocal}. The underlying mechanism of both approaches is the same, i.e., increasing the coupling efficiency between the atom and the pulse modes.

\begin{figure}
    \centering
    \includegraphics[width=9cm]{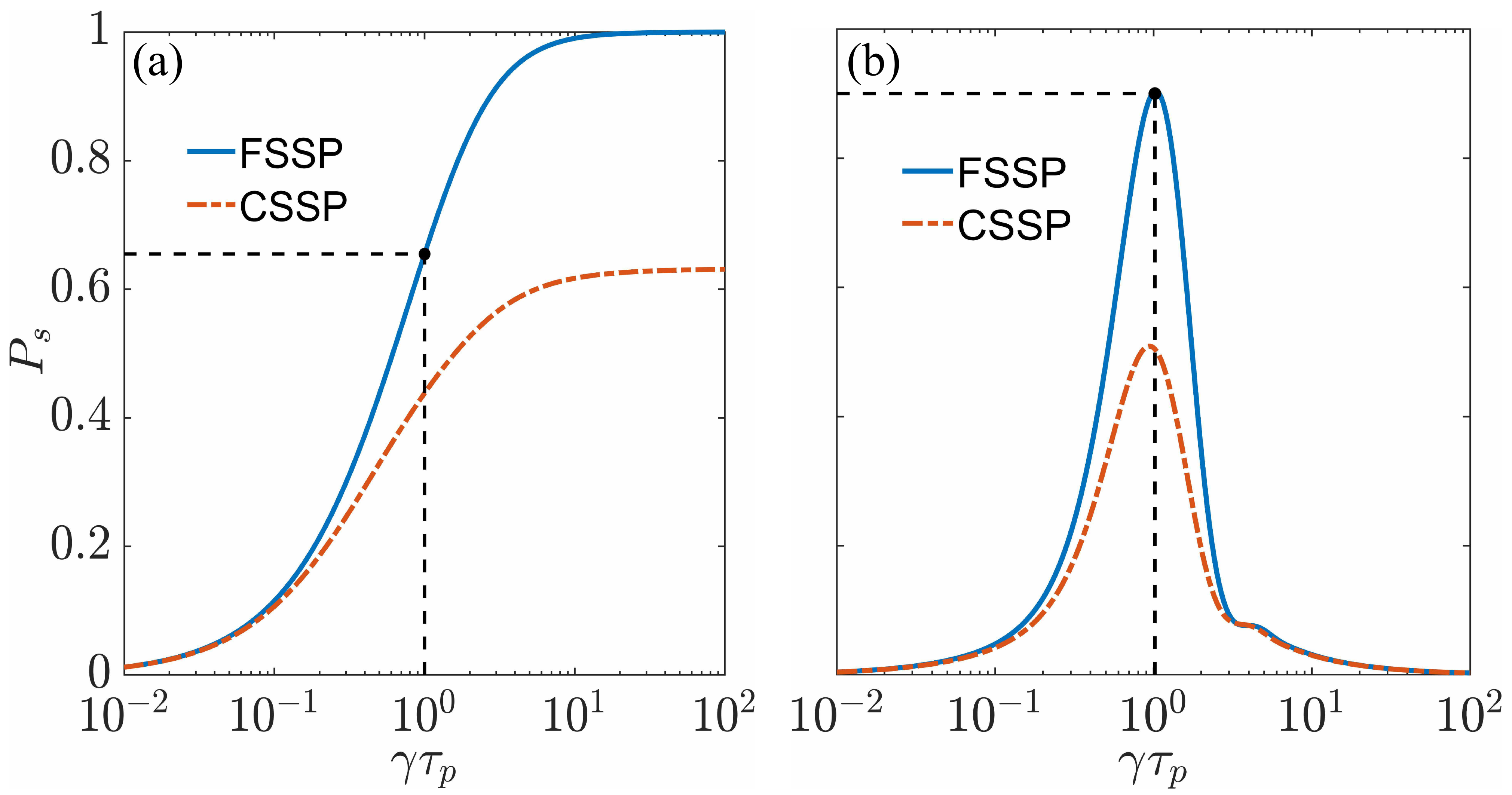}
    \caption{\label{fig:8}Comparison of the improved storage efficiency of Fock-state single-photon (FSSP) and coherent-state single-photon (CSSP) pulses (a) without and (b) with a control pulse. The two-photon detuning is set as $\Delta = 0$. (a) $\gamma_{eg} = \gamma_{es} = \gamma/2$. (b) $\gamma_{eg} = 0.9\gamma$, $\gamma_{es} = 0.1 \gamma$, $\Omega = 0.7\gamma$, $a = 0.9 \tau_p$, and $b = 0.6 \tau_p$.}
\end{figure}

\begin{figure}
    \centering
    \includegraphics[width=8.5cm]{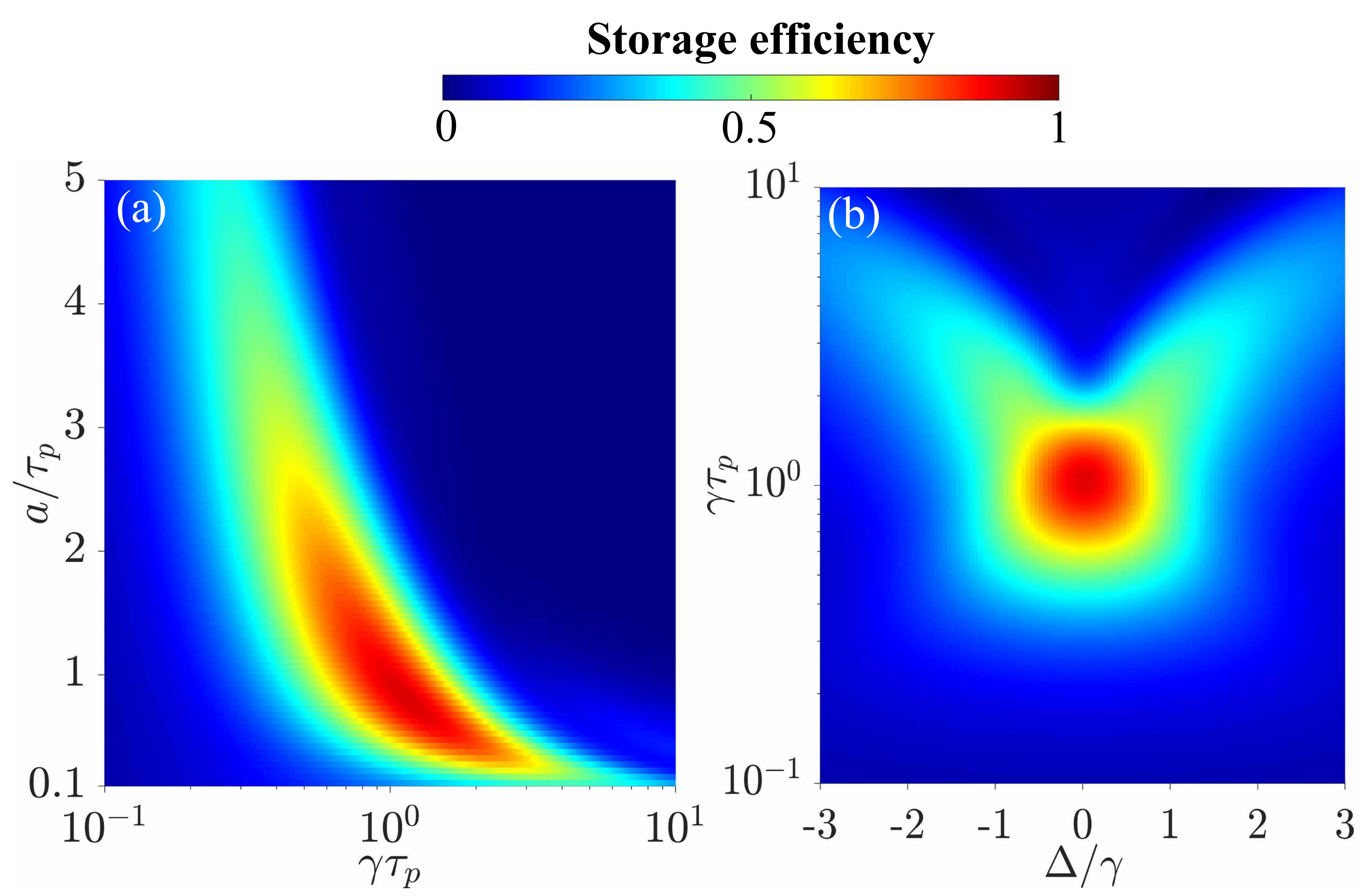}
    \caption{\label{fig:9} Optimization of storage efficiency of a Fock-state single-photon pulse in presence of control field. $\gamma_{eg} = 0.9\gamma$, $\gamma_{es} = 0.1 \gamma$, $\Omega=0.7 \gamma$, $b = 0.6 \tau_p$. (a) Two-photon resonance case with $\Delta = 0$. (b) Off-resonance case with$a = 0.9 \tau_p$.}
    
\end{figure}

For a perfect chiral waveguide, the atom only interacts with photons propagating in one direction [see Fig.~\ref{fig:6}(a)], such as the forward-propagating modes $a(\omega)$. The backward-propagating modes will not contribute to the scattering and storage of the target single-photon pulse. The spontaneous decay of the excited state comes solely from the interaction with forward-propagating modes. In this case, the pumping rate of the single-photon pulse does not change, but the decay rates of state $|e\rangle$ are halved. Thus, the $1/\sqrt{2}$-factor in Eqs.~(\ref{eq:Lp}) and (\ref{eq:MEQrho_01}) will be removed. 

For a Sagnac interferometer case, the incident single-photon pulse will be split into two identical small pulses via a $50:50$ beam splitter. These two small pulses enter the waveguide at two different ends [see Fig.~\ref{fig:6}(b)]. Mathematically, the waveguide modes can always be re-expanded with even and odd modes $a_{\pm}(\omega)=\left[ a(\omega) \pm b(\omega)\right]/\sqrt{2}$. From Eq.~(\ref{eq:Hint}), we see that the atom is only coupled to even modes. Thus, only even modes will contribute to the spontaneous decay of the atomic excited state. By carefully tuning the relative phase between the two small pulses, one can guarantee that the target pulse (i.e., the superposition of two small pulses) only contains even modes. The target pulse is now described by a new single-photon wave-packet creation operator $a_\xi^{\dagger} =\int d\omega \xi (\omega)[a^{\dagger}(\omega)+b^{\dagger}(\omega)]/\sqrt{2}=\int d\omega \xi (\omega)a^{\dagger}_{+}(\omega)$. In this case, the decay rates of state $|e\rangle$ do not change, but the pumping rate of the single-photon pulse gets doubled. Thus, the $1/\sqrt{2}$ factor in Eqs.~(\ref{eq:Lp}) and (\ref{eq:MEQrho_01}) will be removed.   

\begin{figure}
\includegraphics[width=8.5cm]{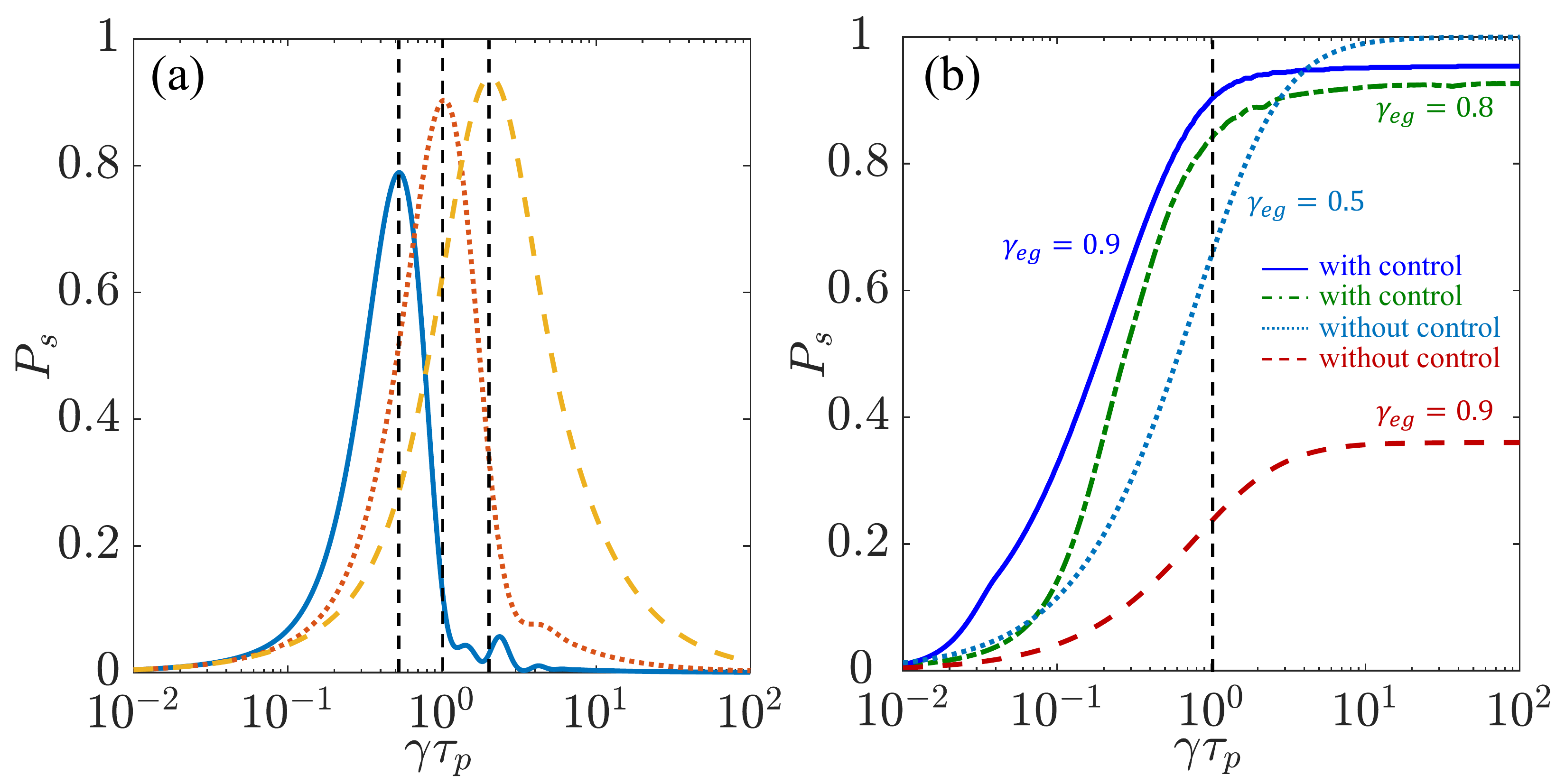}\caption{\label{fig:10} Global optimization of storage efficiency of a Fock-state single-photon pulse in high-dimensional parameter space. (a) The shift of the favorable length of the target pulse. The three lines (blue solid, orange dotted, yellow dashed) are obtained with parameters $b=\{1.3\tau_p,0.6\tau_p,-0.4\tau_p\}$, $a=\{0.7\tau_p,0.9\tau_p,1.2\tau_p\}$, $\Omega=\{1.5\tau_p,0.7\tau_p,0.4\tau_p\}$, and $\Delta = 0$. (b) Comparison of the storage efficiency with and without a control pulse. The blue solid line and green dashed-dotted lines denote the global maximum storage efficiency with $\gamma_{eg} = 0.9\gamma$ and $\gamma_{eg} = 0.8\gamma$, respectively. In the absence of a control pulse, the 
red dashed and navy-blue dotted lines denote the optimal storage efficiency with $\gamma_{eg}=0.9\gamma$ and $\gamma_{eg}=0.5\gamma$, respectively.}
\end{figure} 

We now show that the perfect storage of single-photon pulses in a single three-level atom can be realized with a chiral waveguide or Sagnac interferometer. In Figs.~\ref{fig:7}(a) and ~\ref{fig:7}(b), we plot the storage probability versus $\tau_p$ and $\gamma_{eg}$ for a FSSP and a CSSP pulse, respectively, in the absence of a control pulse. Similar to the regular waveguide case (see Fig.~\ref{fig2}), larger storage efficiency is obtained under the decay-rate matching condition $\gamma_{eg}=\gamma_{es}$. However, the maximum storage efficiency of a FSSP pule can now reach $1$ at the long-pulse limit $\tau_p \gg 1/\gamma$, as shown in Fig.~\ref{fig:8}(a). The upper limit of the storage efficiency of a CSSP pulse has also been raised from $0.4$ to be larger than $0.6$.

The storage process can be accelerated by a control pulse without sacrificing the storage efficiency too much. Similar to Sec.~\ref{sec:control}, there exists \textcolor{blue}{a} favorable pulse length $\tau_p$ in storage efficiency optimization with fixed $b$ and $\Omega$, as shown in Figs.~\ref{fig:8} and ~{\ref{fig:9}}. We emphasize that the maximum storage efficiency in Fig.~\ref{fig:9} is a local one, not the global maximum in the high-dimensional parameter space $\{a,b,\Omega,\Delta,\tau_p\}$. As shown in Fig.~\ref{fig:10}(a), the favorable $\tau_p$ moves toward longer pulses by varying the control pulse parameters, specifically the relative delay $b$. We give the global maximum storage efficiency via brute-force numerical simulations as shown by the blue solid line in Fig.~\ref{fig:10}(b). Compared with cases without a control pulse (the red dashed and navy-blue dotted lines), much larger storage efficiency for relatively short pulses $\tau_p\sim 1/\gamma$ can be obtained under a  control pulse. The storage efficiency of a FSSP pulse with $\tau_p=1/\gamma$ can reach $\approx 0.9$ [see Fig.~\ref{fig:8}(b)]. When the value of $\gamma_{eg}$ is reduced to 0.8$\gamma$, the maximum storage efficiency exhibits a slight decrease, as indicated by the green dashed-dotted line. However, the application of a control pulse does accelerate the storage process. The storage speed is still limited by the total spontaneous decay rate $\gamma$ of the excited state $|e\rangle$.

\section{Conclusion\label{sec:summary}}
We use a simple model, which is composed of a single $\Lambda$-type atom placed in a one-dimensional (1D) waveguide, to explore the limits of single-photon storage. We show that, for a regular waveguide, the storage efficiency of a FSSP pulse is limited to $0.5$ and the efficiency of a CSSP pulse is even lower. Perfect single-photon storage could be achieved by exploiting a chiral waveguide or a Sagnac interferometer. We find that there is a trade-off between storage efficiency and storage speed. A control pulse can be applied to accelerate the storage process. However, the storage speed is ultimately limited by the total decay rate of the excited state involved.

One of the authors (L.P.Y) showed that the absorption speed of a single-photon pulse is limited by the width of the atom-light interaction spectrum~\cite{yang2018concept}. For an atom interacting with 1D waveguide modes, the interaction spectrum is almost flat. Thus, the storage speed is mainly limited by deexcitation processes. In most experiments, an atomic ensemble instead of a single atom was used as the storage media. Due to the photon-induced cooperative dissipation, both super-radiant and subradiant atomic states can be observed~\cite{Asenjo2017exponential}. By carefully selecting the pumping and storage channels, it is possible to achieve high-speed and high-efficiency storage of single-photon pulses. However, storing single-photon pulses in a multiple-atom system is a more intricate task due to the photon-mediated dipole-dipole interaction of the atoms. In our forthcoming work, we will investigate this intriguing problem by examining the scattering of a photon pulse by an atomic ensemble or atomic chain with a control pulse.

\section*{Acknowledgements}
The authors thank Xin Yue and Professor C. P. Sun for the helpful discussion. This work is supported by NSFC Grant
No. 12275048 and National Key R\&D Program of China (Grant No. 2021YFE0193500).

\appendix*

\section{Deduction of Master Equations\label{sec:appendix}}

We give some details of deriving the master equation for an atom driven by a quantum pulse. The Heisenberg equation of a waveguide mode is given by
\begin{equation}
    \dot{a}(\omega,t) = - i \omega a(\omega) - i g_{eg} \sigma_{ge}(t) - i g_{es} \sigma_{se}(t) e^{- i (\omega_c - \omega_0)t},\label{dynamic equation of a}
\end{equation}
where we have set $\hbar = 1$. We integrate the formal solution of $a(\omega,t)$ over $\omega$ to obtain~\cite{gheri1998photon}
\begin{equation}
    \begin{aligned}
        \int a(\omega,t) d\omega &= \sqrt{2 \pi} a_{in}(t) - i \pi g_{eg} \sigma_{ge}(t) - i \pi g_{es} \sigma_{se}(t) e^{- i (\omega_c - \omega_0) t},
    \end{aligned} \label{integrate of a and b}
\end{equation}
where the so-called input-field $a_{in}(t)$ is an explicitly time-dependent operator
\begin{equation}
    \begin{aligned}
        a_{in}(t) &= \frac{1}{\sqrt{2 \pi}} \int_{- \infty}^{\infty} d\omega e^{- i \omega t} a(\omega).
    \end{aligned}
\end{equation}
Note that the two-time commutator of the input field yields a $\delta-$ function
\begin{equation}
    \left[a_{in}(t), a^{\dagger}_{in}(t^{\prime})\right] = \delta(t - t^{\prime}).
\end{equation}
In obtaining Eq.~\eqref{integrate of a and b}, we have used the Wigner-Weisskopf approximation by treating the coupling coefficients as frequency-independent constants
\begin{equation}
        g_{eg}(\omega) \approx g_{eg}(\omega_0)\! =\! \sqrt{\frac{\gamma_{eg}}{4 \pi}},\ 
        g_{es}(\omega) \approx g_{es}(\omega_0) \!= \!\sqrt{\frac{\gamma_{es}}{4 \pi}},
\end{equation}
where $\gamma_{eg}$ and $\gamma_{es}$ are the decay rates of the excited state $|e\rangle$ to ground state $|g\rangle$ and storage state $|s\rangle$, respectively. We can get the similar expression of $b_{in}(t)$ similarly.

The pumping effect from an incident FSSP pulse is characterized by the following relations
\begin{align}
        a_{in}(t) \ket{1_{\rm FS}} \otimes \ket{0_b} & \!=\! \frac{1}{\sqrt{2 \pi}}\! \int\!\! d\omega \xi(\omega) e^{- i \omega t} \ket{0}  \!=\! \tilde{\xi}(t) \ket{0}, \label{eq:ainF} \\
        b_{in}(t) \ket{1_{\rm FS}} \otimes \ket{0_b} & = 0.\label{relation of a_in b_in and psi_R}
\end{align} 
We can obtain the motion equation an arbitrary operator of the system ~\cite{gheri1998photon,Gardiner1994driving}
\begin{widetext}
\begin{equation}
    \begin{aligned}
        \dot{X}(t) = &i [H_s,X(t)] + i \Omega_c(t) [\sigma_{se}^{\dagger}(t) + \sigma_{se}(t),X(t)] \\
        &+ [\sigma_{ge}^{\dagger}(t),X(t)] \left[i \sqrt{\frac{\gamma_{eg}}{2}} a_{in}(t) + i \sqrt{\frac{\gamma_{eg}}{2}} b_{in}(t) + \frac{\gamma_{eg}}{2} \sigma_{ge}(t) + \frac{\sqrt{\gamma_{eg} \gamma_{es}}}{2} \sigma_{se}(t) e^{- i (\omega_c - \omega_0) t}\right] \\
        &+ e^{i (\omega_c - \omega_0)t} [\sigma_{se}^{\dagger}(t),X(t)] \left[i \sqrt{\frac{\gamma_{es}}{2}} a_{in}(t) + i \sqrt{\frac{\gamma_{es}}{2}} b_{in}(t) + \frac{\sqrt{\gamma_{eg} \gamma_{es}}}{2} \sigma_{ge}(t) + \frac{\gamma_{es}}{2} \sigma_{se}(t) e^{- i (\omega_c - \omega_0) t}\right]\\
        &+ \left[i \sqrt{\frac{\gamma_{eg}}{2}} a_{in}^{\dagger}(t) + i \sqrt{\frac{\gamma_{eg}}{2}} b_{in}^{\dagger}(t) - \frac{\gamma_{eg}}{2} \sigma_{ge}^{\dagger}(t) - \frac{\sqrt{\gamma_{eg} \gamma_{es}}}{2} \sigma_{se}^{\dagger}(t) e^{i (\omega_c - \omega_0) t}\right] [\sigma_{ge}(t),X(t)]\\ 
        &+ e^{- i (\omega_c - \omega_0)t} \left[ i \sqrt{\frac{\gamma_{es}}{2}} a_{in}^{\dagger}(t) + i \sqrt{\frac{\gamma_{es}}{2}} b_{in}^{\dagger}(t) - \frac{\sqrt{\gamma_{eg} \gamma_{es}}}{2} \sigma_{ge}^{\dagger}(t) - \frac{\gamma_{es}}{2} \sigma_{se}^{\dagger}(t) e^{i (\omega_c - \omega_0) t}\right] [\sigma_{se}(t),X(t)]. \label{equation of X(t)}
    \end{aligned}
\end{equation}
\end{widetext}
Using the relations \eqref{relation of a_in b_in and psi_R} and \eqref{equation of X(t)}, we obtain the motion equations for $\rho$, $\rho_{01}$, and $\rho_{00}$ in the main text [i.e., Eqs.~(\ref{evolution equation of rhoS})-(\ref{evolution equation of rhor})]. Note that the fast-oscillating terms have been neglected. %Hence, we have a series of closed equations for the density matrices. Because of the homogeneity of $\rho_{00}(t)$ and initial value is $\rho_{00}(0) = \ket{g}\bra{g}$, we can get the solution of $\rho_{00}(t)$ as $\rho_{00}(t)= \ket{g}\bra{g}$. Therefore, the evolution equation of $\rho_{01}(t)$ can be simplified as
% \begin{equation}
%     \dot{\rho}_{01}(t) = \mathcal{L} \rho_{01}(t) + i \sqrt{\frac{\gamma_{eg}}{2}} \tilde{\xi}^*(t) \sigma_{ge},
% \end{equation}
% which makes the final master equation easier.
Different from Eq.~(\ref{eq:ainF}) for a FSSP pulse, the action of the input-field operator on a CSSP pulse is given by
\begin{equation}
a_{in}(t) \ket{1_{\rm{CS}}}\otimes \ket{0_b} =  \tilde{\xi}(t) \ket{1_{\rm{CS}}}\otimes \ket{0_b}. \end{equation}
In this case, we obtain a single master equation as given in Eq.~(\ref{eq:Lp}), where the CSSP pulse functions as a classical pump.

\bibliography{main}
\end{document}